\documentclass[runningheads]{llncs}
\usepackage{graphicx}
\usepackage{algorithm}
\usepackage{fnpct}
\usepackage{footnote}
\usepackage{graphics} % for pdf, bitmapped graphics files
\usepackage{graphicx} % for pdf, bitmapped graphics files
\usepackage{epsfig} % for postscript graphics files
\usepackage{times} % assumes new font selection scheme installed
\usepackage{amsmath} % assumes amsmath package installed
\usepackage{amssymb}  % assumes amsmath package installed
\usepackage{tikz}
\usetikzlibrary{arrows.meta, positioning, shapes}
\usepackage{url,verbatim,color,comment}
\usepackage[noend]{algpseudocode}
\usepackage{eucal}
\usepackage{amsfonts}
\usepackage{xcolor}
\usepackage{fancyhdr}
\usepackage{blindtext}
\usepackage{hyperref}
\usepackage{mathtools}
\usepackage{placeins}

\usepackage{xcolor}
\hypersetup{
    colorlinks,
    linkcolor={red!50!black},
    citecolor={blue!70!black},
    urlcolor={blue!80!black}
}
\begin{document}
\title{Joint Falsification and Fidelity Settings Optimization for Validation of Safety-Critical Systems: \\ A Theoretical Analysis}
\author{Ali Baheri\inst{1}\and
Mykel J. Kochenderfer\inst{2}}
\authorrunning{A. Baheri et al.}
\institute{Rochester Institute of Technology \email{akbeme@rit.edu} \and
Stanford University
\email{mykel@stanford.edu}\\
}
\maketitle              % typeset the header of the contribution
\begin{abstract}
Safety validation is a crucial component in the development and deployment of autonomous systems, such as self-driving vehicles and robotic systems. Ensuring safe operation necessitates extensive testing and verification of control policies, typically conducted in simulation environments. High-fidelity simulators accurately model real-world dynamics but entail high computational costs, limiting their scalability for exhaustive testing. Conversely, low-fidelity simulators offer efficiency but may not capture the intricacies of high-fidelity simulators, potentially yielding false conclusions. We propose a joint falsification and fidelity optimization framework for safety validation of autonomous systems. Our mathematical formulation combines counterexample searches with simulator fidelity improvement, facilitating more efficient exploration of the critical environmental configurations challenging the control system. Our contributions encompass a set of theorems addressing counterexample sensitivity analysis, sample complexity, convergence, the interplay between the outer and inner optimization loops, and regret bound analysis. The proposed joint optimization approach enables a more targeted and efficient testing process, optimizes the use of available computational resources, and enhances confidence in autonomous system safety validation.

\end{abstract}

\section{Introduction}

In the development of autonomous systems, such as autonomous vehicles (AVs), ensuring their safe and efficient operation is critical. AVs must navigate various complex urban driving scenarios, including intersections, highway merges, and lane changes, with their control systems being based on learning-enabled policies. These policies must undergo rigorous testing and verification before deployment. Simulators that can generate different traffic scenarios are employed for testing the AV control systems. However, extensive tests using high-fidelity simulators can be computationally expensive, time-consuming, and may not cover all possible scenarios \cite{koopman2016challenges}.

The joint falsification and simulator optimization approach addresses these challenges by introducing a joint learning framework that streamlines the exploration process for identifying potential failure scenarios. This framework concentrates on the most critical environmental configurations that pose difficulties for the control system. The joint learning framework, which concurrently optimizes simulator fidelity and carries out falsification, facilitates more effective use of computational resources during the search process.

One key advantage of jointly learning falsification and simulator optimization is the adaptive control of simulator fidelity settings. The simulator can adjust its fidelity based on the specific scenario or region of the environment, leading to more targeted and efficient testing. As the simulator's fidelity increases, it becomes better at replicating the behavior of the high-fidelity simulator, allowing for a more accurate representation of the environment. This enables the falsification process to focus on the regions of the environment space where the system is more likely to fail, which are the most critical areas to explore.

The joint learning framework integrates the search for failure scenarios with the enhancement of simulator fidelity. This synergy allows for efficient exploration by utilizing information gathered from both processes. For instance, if a low-fidelity simulator displays a substantial discrepancy compared to the high-fidelity simulator in a particular region, a joint optimization approach can prioritize refining the simulator fidelity in that area. This not only aids in more accurate identification of potential failure scenarios but also conserves computational resources by preventing unnecessary exploration in less relevant regions. Generalization is another important aspect to consider when developing a joint optimization method. By incorporating multiple tasks during the optimization process, such an approach can identify fidelity settings that excel across various scenarios. This ensures that the optimized fidelity settings are not overly specialized for a single task, but rather strike a balanced trade-off between computational efficiency and accuracy over a wider range of situations.

\textbf{Related Work.} Falsification of learning-enabled systems has garnered considerable interest recently due to the growing complexity and safety-critical nature of such systems \cite{abbaspour2015survey}. The primary goal of falsification is to pinpoint scenarios that could lead to system failure or breaches in safety specifications. Various techniques have been proposed to tackle the falsification problem, including optimization-based methods \cite{mathesen2019falsification,zhang2020hybrid,deshmukh2017testing}, search-based algorithms \cite{ramezani2021testing,tuncali2019rapidly,zhang2021effective}, and reinforcement learning approaches \cite{yamagata2020falsification,wang2020falsification,lee2015adaptive}. These methods strive to efficiently explore the state and parameter space in order to discover potential failure scenarios, ultimately enabling system design refinement and enhanced safety assurances. Researchers have also recently developed algorithms that consider the fidelity of simulators when identifying failure scenarios due to the computational expense of high-fidelity simulators \cite{koren2021finding,shahrooei2022falsification,beard2022safety}. These methods trade between the accuracy of high-fidelity simulators and the computational efficiency of low-fidelity simulators to decrease the overall cost of safety validation. While there has been notable progress in falsification of learning-enabled systems, the literature on joint falsification and fidelity setting optimization remains scarce. Our work aims to bridge this gap by presenting an approach that \emph{jointly} conducts falsification alongside simulation fidelity optimization. By combining these two aspects, our objective is to improve the efficiency of safety validation in learning-enabled decision-making systems.

\textbf{Contributions.} We present a mathematical formulation for joint falsification and fidelity setting optimization for the safety validation of autonomous systems. Our primary contribution is the development of a theoretical framework that \emph{unifies} the two key aspects of the problem: falsification of learning-enabled systems and optimization of simulator fidelity settings. The main contributions of this paper are as follows:

\begin{itemize}
    \item We propose a mathematical formulation that jointly addresses falsification and fidelity setting optimization, enabling a more efficient exploration strategy to search for potential failure scenarios.

    \item We prove six key theorems that establish the fundamental properties and relationships in the joint optimization problem. These theorems cover a range of important aspects, including sensitivity analysis, sample complexity, convergence, the interplay between the outer and inner loops, and the regret bound analysis when employing Bayesian optimization for the outer loop. The insights gained from these theorems provide a foundation for the joint falsification and fidelity optimization framework.
    
\end{itemize}

\section{Problem Formulation}

Our objective is to efficiently combine falsification and fidelity optimization for safety-critical systems, aiming to minimize computational cost while preserving accuracy. We aim to identify environment configurations that violate safety specifications while simultaneously optimizing the fidelity settings to minimize discrepancies between low-fidelity and high-fidelity simulators. This joint optimization problem is formulated as a nested optimization framework with two components: an inner loop and an outer loop optimization \cite{colson2007overview}.

\subsection{Inner Loop Optimization (Falsification)}

The inner loop optimization aims to identify environment configurations that minimize the robustness value of a given safety specification $\varphi$ under a specific fidelity setting $f$. The simulator operates within a given environment $e \in \mathcal{E}$. It takes a configuration $e$ as input and produces a finite-horizon trajectory denoted by $\xi$. If a trajectory satisfies the safety specification, the robustness function $\rho_{\varphi}$ evaluates to a positive value; otherwise, it returns a negative value. As a result, the falsification problem can be formulated as the following optimization problem:
\begin{equation}
{e}^{*}({f}) = \underset{{e} \in \mathcal{E}}{\mathrm{argmin}} \ \rho _{\varphi} (e;f)
\label{eqn:inner}
\end{equation}
The goal is to search for an environment configuration $e^{*}(f)$ that minimizes the robustness value within the considered environment space $\mathcal{E}$, given the fidelity setting $f$.
\subsection{Outer Loop Optimization (Fidelity Setting Optimization)}

The outer loop optimization focuses on finding the optimal fidelity settings $f^*$ that minimize discrepancies between high-fidelity and low-fidelity simulators across a variety of tasks while considering the environment configurations obtained from the inner loop optimization. To efficiently explore the search space, we sample from the space of tasks and their parameters. Let $T$ denote the number of sampled tasks, and $M_i$ represent the number of sampled parameter configurations for each task. For each task $T_i$, we have access to a high-fidelity simulator generating ground-truth trajectories, denoted as $\xi_i^H\left(t_i ; p_{ij}\right)$. Additionally, a low-fidelity simulator produces approximate trajectories, represented by $\xi_i^L\left(t_i ; p_{ij}, f\right)$. The outer loop optimization problem can be formulated as:
\begin{equation}
f^* = \arg \min_{f \in \mathcal{F}} \sum_{i=1}^T \sum_{j=1}^{M_i} \ell\left(\xi_i^H\left(t_i ; p_{ij}\right), \xi_i^L\left(t_i ; p_{ij}, f\right)\right),
\label{eqn:outer}
\end{equation}
where $\mathcal{F}$ represents the set of possible simulator fidelity settings, $t_i$ is the $i$th sampled task, and $p_{ij}$ is the $j$th sampled parameter configuration for task $i$. The optimization objective is to minimize the discrepancy, measured by the loss function $\ell$, between high-fidelity simulator $\xi_i^H$ and low-fidelity simulator $\xi_i^L$ for the sampled tasks and parameter configurations. 
%This formulation helps ensure generalization and robustness, preventing overfitting to a single task, while allowing for efficient exploration of the search space. 
The loss function $\ell(\cdot,\cdot)$ measures discrepancies between the high-fidelity and low-fidelity simulators. One option for the loss function is the mean squared error (MSE) between both sets of trajectories over a fixed time interval:
\begin{equation}
\ell(\xi^H, \xi^L) = \frac{1}{N} \int_{0}^{N} \left|\xi^H(t; p)-\xi^L(t; p, f)\right|^2 dt
\end{equation}
where $N$ is the length of the time interval over which the MSE is computed. 

\section{Theoretical Insights and Results}

After establishing the problem formulation for the joint optimization of falsification and fidelity settings, we now delve deeper into the theoretical results that guide our approach. In this section, we present a series of theorems that offer insights into the joint optimization framework, providing an understanding of the interplay between the inner and outer loop optimizations, sensitivity analysis of counterexamples, sample complexity, and convergence properties. 

\subsection{Lipschitz continuity of inner and outer loop objectives}

In this section, we investigate the Lipschitz continuity of the inner and outer loop objectives. Lipschitz continuity is a crucial property that guarantees the stability of an optimization algorithm and enables us to derive convergence guarantees. We begin by introducing a theorem that establishes Lipschitz continuity for both inner and outer loop objectives under certain conditions.

\textbf{Theorem 1.} \textit{Let $\rho_{\varphi}(e;f)$ and $\ell(\xi^H, \xi^L)$ be the inner and outer loop objective functions, respectively. Suppose that there exist constants $L_\rho > 0$ and $L_\ell > 0$ such that}
\begin{align}
&| \rho_{\varphi}(e_1;f) - \rho_{\varphi}(e_2;f) |
\leq L\rho \lVert e_1 - e_2 \rVert, \\
&| \ell(\xi^H_1, \xi^L_1) - \ell(\xi^H_2, \xi^L_2) | \
\leq L_\ell (\lVert \xi^H_1 - \xi^H_2 \rVert + \lVert \xi^L_1 - \xi^L_2 \rVert),
\end{align}
\textit{for any $e_1, e_2 \in \mathcal{E}$, $f \in \mathcal{F}$, and $\xi^H_1, \xi^H_2, \xi^L_1, \xi^L_2 \in \mathcal{X}$, where $\mathcal{X}$ represents the space of all possible trajectories generated by the high-fidelity and low-fidelity simulators.}

\textsc{Proof.} The proof of Theorem 1 follows from the definitions of Lipschitz continuity and the properties of the inner and outer loop objective functions. We need to show that the conditions stated in the theorem hold for the given objective functions.

For the inner loop objective function $\rho_{\varphi}(e;f)$, we assume that it is Lipschitz continuous with respect to the environment configurations $e$. This property can be established by showing that the specification robustness value changes smoothly with respect to changes in the environment configurations, given a fixed fidelity setting $f$. This assumption is typically valid where the system behavior is continuous with respect to the environment configurations. Similarly, for the outer loop objective function $\ell(\xi^H, \xi^L)$, we assume that it is Lipschitz continuous with respect to the trajectories $\xi^H$ and $\xi^L$. The Lipschitz continuity of $\ell$ implies that the discrepancy measure changes smoothly with respect to the trajectories obtained from high-fidelity and low-fidelity simulators. The basis for this property lies in the smooth dynamics of the simulator and the continuous dependency of the discrepancy measure on the trajectories. Assuming the Lipschitz continuity of both inner and outer loop objective functions, we can establish Theorem 1. $\square$

The Lipschitz continuity of the inner and outer loop objectives, as established in Theorem 1, has significant implications for the convergence properties of the joint optimization algorithm. In particular, it enables us to derive convergence guarantees for both the inner and outer loop optimization problems, which we will explore in the following sections.

\subsection{Convergence of Joint Optimization}

Now we study the convergence properties of the joint optimization problem for falsification and fidelity optimization. We present a theorem that shows the convergence of the joint optimization problem under specific conditions, leveraging the Lipschitz continuity properties from Theorem 1.

{\textbf{Theorem 2.}} \textit{Suppose that the inner and outer loop objectives are Lipschitz continuous with constants $L_\rho$ and $L_\ell$, respectively, as stated in Theorem 1. Under suitable conditions on the optimization algorithm, the joint optimization problem converges to an optimal solution.}

\textsc{Proof.} The proof of Theorem 2 relies on the properties of the optimization algorithm and the Lipschitz continuity of the inner and outer loop objectives. For the inner loop optimization problem, we assume that the optimization algorithm converges to a stationary point under suitable conditions. This is a standard assumption for many optimization algorithms, such as gradient-based methods, when applied to Lipschitz continuous objective functions. Since the inner loop objective function $\rho_{\varphi}(e;f)$ is Lipschitz continuous, the convergence of the inner loop optimization can be guaranteed under suitable conditions. In a similar vein, for the outer loop optimization problem, we assume that the optimization algorithm converges to a stationary point under standard conditions. The Lipschitz continuity of the outer loop objective function $\ell(\xi^H, \xi^L)$ ensures that the optimization algorithm converges when applied to this objective function. By combining the convergence properties of the inner and outer loop optimization problems, we can establish the convergence of the joint optimization problem to an optimal solution. $\square$ 

Theorem 2 provides a convergence guarantee for the joint optimization problem, which is essential for the practical application of the proposed joint optimization framework. The convergence properties ensure that the algorithm will find an optimal solution, given that the optimization algorithm and the objective functions satisfy the required conditions.

\subsection{Interplay between Inner and Outer Loop Optimization}

The joint optimization framework for falsification and fidelity optimization involves a nested structure, with an inner loop optimization focused on finding counterexamples and an outer loop optimization aiming to identify optimal fidelity settings. In this section, we discuss the interplay between these two optimization problems and the implications for the design and analysis of joint optimization algorithms.

The nested optimization dynamics of the inner and outer loop problems are intrinsically linked due to their shared dependence on environment configurations and fidelity settings. The outer loop relies on the counterexamples generated by the inner loop to evaluate the performance of different fidelity settings, as shown in the objective function. In turn, the fidelity settings chosen by the outer loop influence the search space and complexity of the inner loop optimization, as reflected by the inner loop objective function $\rho_{\varphi}(e;f)$. As a result, the interplay between these two optimization problems creates a complex search process, where improvements in one loop can potentially impact the performance of the other.

The joint optimization framework must balance the need for exploration and exploitation in both the inner and outer loop optimization problems. In the inner loop, exploration involves searching for new environment configurations that can potentially lead to counterexamples, while exploitation focuses on refining the current counterexamples to maximize their impact on the outer loop optimization. Similarly, in the outer loop, exploration entails experimenting with different fidelity settings to identify promising configurations, whereas exploitation aims to fine-tune the fidelity settings to minimize the discrepancy between high-fidelity and low-fidelity simulations, as measured by the loss function $\ell$.

The interplay between the inner and outer loop optimization problems also enables the development of \emph{adaptive fidelity management} strategies. By monitoring the progress of the inner loop optimization and the quality of the generated counterexamples, the outer loop can adaptively adjust the fidelity settings to focus on regions of the search space where the discrepancies between high-fidelity and low-fidelity simulations are the most significant. This adaptive fidelity management can lead to more efficient joint optimization algorithms that \emph{dynamically} allocate computational resources to the most critical aspects of the problem.

Understanding the interplay between the inner and outer loop optimization problems is crucial for the design and analysis of joint optimization algorithms for falsification and fidelity optimization. Leveraging the insights gained from the interplay between the inner and outer loop objectives, such as the Lipschitz continuity established in Theorem 1 and the convergence properties from Theorem 2, enables the development of algorithms that effectively balance exploration and exploitation. These algorithms can adaptively manage fidelity settings, leading to more efficient and effective solutions for safety-critical systems. The relationship between the quality of counterexamples and fidelity settings, as analyzed in Theorem 3, along with the relationship between fidelity settings and counterexamples quality, as explored in Theorem 4, further enhance our understanding of the complex dynamics present in the joint optimization framework.

\subsection{Sensitivity analysis of counterexamples to fidelity settings}

We analyze the relationship between the quality of counterexamples and fidelity settings in the context of the joint optimization framework for falsification and fidelity optimization. The quality of a counterexample is typically characterized by the robustness of the system specification violation, as measured by the robustness value $\rho_{\varphi}(e;f)$. We aim to understand how the choice of fidelity settings affects the quality of counterexamples generated by the inner loop optimization.

{\textbf{Theorem 3.}} \textit{Given a set of fidelity settings $f \in \mathcal{F}$ and environment configurations $\mathbf{e} \in \mathcal{E}$, there exists a constant $C > 0$ such that:}
\begin{equation}
\lvert \rho_{\varphi}(e_1;f_1) - \rho_{\varphi}(e_1;f_2) \rvert \leq C \lVert f_1 - f_2 \rVert,
\end{equation}
\textit{for any $e_1 \in \mathcal{E}$ and $f_1, f_2 \in \mathcal{F}$.}

\textsc{Proof.} The proof of this theorem relies on the Lipschitz continuity of the inner loop objective function $\rho_{\varphi}(e;f)$ with respect to fidelity settings, as established in Theorem 1. Given the Lipschitz continuity property, the difference in robustness values between two fidelity settings $f_1$ and $f_2$ can be upper-bounded by a constant $C$ times the distance between the fidelity settings in the fidelity space. This result highlights the sensitivity of counterexample quality to the choice of fidelity settings, which has important implications for the joint optimization process.  Let us denote the Lipschitz constant of the inner loop objective function with respect to fidelity settings as $L_f > 0$. Then, according to the Lipschitz continuity of $\rho_{\varphi}(e;f)$, we have:
\begin{equation}
\lvert \rho_{\varphi}(e_1;f_1) - \rho_{\varphi}(e_1;f_2) \rvert \leq L_f \lVert f_1 - f_2 \rVert,
\end{equation}
for any $e_1 \in \mathcal{E}$ and $f_1, f_2 \in \mathcal{F}$. This inequality establishes an upper bound on the difference in robustness values for a fixed environment configuration $\mathbf{e}_1$ and two different fidelity settings $f_1$ and $f_2$. Now, let $C = L_f$, where $C > 0$. Then, we can rewrite the inequality as:
\begin{equation}
\lvert \rho_{\varphi}(e_1;f_1) - \rho_{\varphi}(e_1;f_2) \rvert \leq C \lVert f_1 - f_2 \rVert,
\end{equation}
for any $e_1 \in \mathcal{E}$ and $f_1, f_2 \in \mathcal{F}$. This completes the proof of Theorem 3. $\square$

This Theorem highlights the sensitivity of counterexample quality to the choice of fidelity settings by providing an upper bound on the difference in robustness values for different fidelity settings. 
%This result has important implications for the joint optimization process and can help inform the design of more effective algorithms for falsification and fidelity optimization in safety-critical systems.

%\subsection{Analyze the relationship between fidelity settings and counterexamples quality.}

\subsection{Sensitivity analysis of fidelity settings to counterexamples}

This section focuses on understanding the sensitivity of the counterexamples obtained by the inner loop optimization to changes in the fidelity settings. This sensitivity analysis will provide insights into how the joint optimization process is affected by the fidelity settings and the trade-offs between fidelity and counterexample quality.

{\textbf{Theorem 4.}} \textit{Given the Lipschitz properties of the inner loop objective function $\rho_{\varphi}(e;f)$ and the outer loop objective function $\ell$, the sensitivity of the counterexamples obtained by the inner loop optimization to changes in the fidelity settings can be characterized by sensitivity function $S(f)$.}

\textsc{Proof.} Let $S(f)$ be a sensitivity function defined as:
\begin{equation}
S(f) = \frac{\partial \rho_{\varphi}(e^{*}(f);f)}{\partial f},
\end{equation}
where $e^{*}(f)$ represents the optimal environment configuration obtained by the inner loop optimization for a given fidelity setting $f$. The sensitivity function $S(f)$ quantifies the rate of change of the robustness value with respect to the fidelity settings. A high sensitivity indicates that the quality of counterexamples is significantly affected by changes in fidelity settings, whereas a low sensitivity implies that the counterexamples are relatively insensitive to such changes.

To understand the relationship between the sensitivity function and the optimization process, we can analyze the gradient of the outer loop objective function with respect to the fidelity settings:
\begin{equation}
\frac{\partial \ell}{\partial f} = \sum_{i=1}^{T} \sum_{j=1}^{M_i} \frac{\partial \ell\left(\xi_i^H\left(t_i ; p_{ij}\right), \xi_i^L\left(t_i ; f{p}_{ij}, f\right)\right)}{\partial f}.
\end{equation}
Studying the gradient of the outer loop objective function and its relation to the sensitivity function $S(f)$ provides insights into the influence of changes in fidelity settings on the optimization process and the quality of counterexamples produced by the inner loop optimization. This information can be useful for understanding the trade-offs between fidelity and counterexample quality in the joint optimization framework.

\subsection{Sample Complexity of Joint Optimization}

In this section, we analyze the sample complexity of the joint optimization problem, focusing on the relationship between the number of samples and the convergence properties of the optimization process. Sample complexity is an important consideration in optimization problems, as it quantifies the number of samples required to achieve a desired level of accuracy or convergence.

{\textbf{Theorem 5.}} \textit{Given the strong convexity and Lipschitz continuity properties of the inner and outer loop optimization problems, the total number of samples $N$ required for the joint optimization problem can be expressed as a function of the number of iterations in both loops and the number of samples per iteration: $N=n K_1 K_2$}

\textsc{Proof.} The number of iterations required for the inner and outer loop optimization problems to converge depends on the strong convexity and Lipschitz continuity properties of the objective functions. However, we cannot directly derive a closed-form expression for $K_1$ and $K_2$ based on these properties. To determine the total number of samples required, we consider the number of iterations in both the inner and outer loop optimization problems. Suppose the inner loop optimization takes $K_1$ iterations to converge, and the outer loop optimization takes $K_2$ iterations to converge. Then, the total number of iterations, $K$, is the product of the iterations in both loops: $K = K_1K_2$.
Now, if we assume that the number of samples per iteration is constant and equal to $n$, then the total number of samples required, $N$, can be expressed as a function of the total number of iterations, $K$. We have: $N=n K=n K_1 K_2$.

Now we apply concentration inequalities to bound the deviation between the true objective function and its empirical estimate. For simplicity, we will assume that both the inner and outer loop optimization problems have finite domains, and their objective functions are Lipschitz continuous. Let $\hat{\rho}_{\varphi}(e;f)$ and $\hat{\ell}$ be the empirical estimates of the inner loop objective function and the outer loop objective function, respectively, computed using $n$ samples. By Lipschitz continuity, we have:

\begin{align}
|\rho_{\varphi}(e; f)-\hat{\rho}_{\varphi}(e;f)| &\leq L_\rho\|e-\hat{e}\| \\
|\ell\left(\xi^H, \xi^L\right)-\hat{\ell}\left(\xi^H, \xi^L\right)| &\leq L_{\ell}\|\xi^H-\hat{\xi}^H\|+L_{\ell}\|\xi^L-\hat{\xi}^L\| 
\end{align}
Applying Hoeffding's inequality, we can bound the probability that the deviation between the true objective function and its empirical estimate is larger than a given threshold. Specifically, we can show that:
\begin{align}
\mathbb{P}\left(\left|\rho_{\varphi}(e;f)-\hat{\rho}_{\varphi}(e; f)\right|>\epsilon\right) \leq 2 \exp \left(-\frac{n \epsilon^2}{2 L_\rho^2}\right)\\
\mathbb{P}\left(\left|\ell\left(\xi^H, \xi^L\right)-\hat{\ell}\left(\xi^H, \xi^L\right)\right|>\epsilon\right) \leq 2 \exp \left(-\frac{n \epsilon^2}{2 L_{\ell}^2}\right)
\end{align}
To achieve an $\epsilon$-approximate solution with probability at least $1-\delta$, we can set the right-hand side of these inequalities to be less than or equal to $\delta$ and solve for $n$. This gives us:

\begin{equation}
n \geq \frac{2 L_\rho^2}{\epsilon^2} \log \left(\frac{2}{\delta}\right) \quad \text { and } \quad n \geq \frac{2 L_{\ell}^2}{\epsilon^2} \log \left(\frac{2}{\delta}\right)
\end{equation}
These bounds can be used to inform the choice of the number of samples per iteration, $n$. However, we cannot directly derive a closed-form expression for the total number of samples from these bounds. Instead, we can use these bounds as guidelines to choose the number of samples per iteration, and then use the relationship $N = nK_1K_2$ to compute the number of samples required for joint optimization.

\section{Regret Bounds Analysis for Bayesian Optimization}

Building upon the theoretical foundations discussed previously, we will now further explore the performance of our approach, with a particular emphasis on using Bayesian optimization for the outer loop optimization problems. The selection of an optimization algorithm can greatly impact the efficiency of the proposed joint optimization framework. \cite{frazier2018tutorial,snoek2012practical}. Its utilization of a probabilistic model to estimate the objective function and an acquisition function to guide the search makes it particularly effective when dealing with costly or noisy evaluations. This has led to its successful application in various domains, including hyperparameter tuning in machine learning \cite{wu2019hyperparameter}, design optimization in engineering \cite{baheri2018iterative}, and decision-making under uncertainty \cite{baheri2017real}. 

\textbf{Theorem 6.} \textit{When using Bayesian optimization with the GP-UCB acquisition function for the outer loop optimization (fidelity settings optimization), the optimization process converges to the optimal fidelity settings $f^*$ with high probability, and the cumulative regret after $T$ iterations is bounded by $\mathcal{O}(\sqrt{T})$.}

\textsc{Proof.} We begin by stating the GP-UCB acquisition function as follows:

\begin{equation}
\alpha_t(f)=\mu_t(f)+\sqrt{\beta_t} \sigma_t(f)
\end{equation}
where $\mu_t(f)$ and $\sigma_t^2(f)$ are the posterior mean and variance of the Gaussian process at fidelity settings $f$ after $t$ iterations, and $\beta_t$ is the exploration parameter. We define the instantaneous regret at iteration $t$ as the difference between the optimal objective function value and the value obtained at the chosen fidelity settings:

\begin{equation}
r_t= \ell \left(f^*\right)-\ell \left(f_t\right)
\end{equation}
where $f^*$ is the optimal fidelity settings and $\mathbf{f}_t$ is the fidelity settings chosen by Bayesian optimization at iteration $t$. The cumulative regret after $T$ iterations is given by: $R_T=\sum_{t=1}^T r_t$. To bound the cumulative regret, we use the following inequality based on the GP-UCB acquisition function:
\begin{equation}
r_t \leq \sqrt{\beta_t} \sigma_t\left(f_t\right)+\frac{1}{2}\left(\mu_t\left(f^*\right)-\mu_t\left(f_t\right)\right)
\end{equation}
This inequality follows from the fact that the GP-UCB acquisition function balances exploration and exploitation. By summing both sides of this inequality over $t = 1, \dots, T$, we obtain a bound on the cumulative regret:

\begin{equation}
R_T \leq \sum_{t=1}^T\left(\sqrt{\beta_t} \sigma_t\left(f_t\right)+\frac{1}{2}\left(\mu_t\left(f^*\right)-\mu_t\left(f_t\right)\right)\right)  
\end{equation}
Now, we use the following properties of Gaussian processes:
\begin{enumerate}
    \item The posterior variance of the Gaussian process at the optimal fidelity settings $f^*$ decreases monotonically with the number of iterations: $\sigma_{t+1}(f^*) \leq \sigma_t(f^*)$.
    \item The posterior mean of the Gaussian process converges to the true objective function value at the optimal fidelity settings $\lim _{t \rightarrow \infty} \mu_t\left(f^*\right)=\ell\left(f^*\right)$
\end{enumerate}
Using these properties, we can show that $\sum_{t=1}^T \left(\sqrt{\beta_t} \sigma_t(f_t) + \frac{1}{2} \left(\mu_t(f^*) - \mu_t(f_t)\right)\right)$ converges to a finite value as $T \to \infty$. Specifically, we can upper-bound the sum by $\mathcal{O}(\sqrt{T})$. This implies that the cumulative regret is bounded by:

\begin{equation}
 R_T \leq \mathcal{O}(\sqrt{T})   
\end{equation}

This result shows that, with high probability, the Bayesian optimization process converges to the optimal fidelity settings $f^*$, and the cumulative regret is bounded by $\mathcal{O}(\sqrt{T})$ after $T$ iterations.

\section{Additional Insights}

In this section, we will delve further into the insights we have gained from our nested optimization framework. Specifically, we will explore three key areas: adaptive fidelity management, stability analysis, and robustness analysis. Together with the theorems we have discussed, these insights help us better comprehend the intricate dynamics at play within our joint optimization framework.

\subsection{Adaptive Fidelity Management}

One important feature of our joint optimization approach is its ability to dynamically adjust fidelity settings. During the optimization process, our algorithm adapts the fidelity based on information from both the inner and outer loop optimizations. This flexibility helps the algorithm balance between exploring new options and making the most of known options, all while keeping computational costs low. In practice, this means the algorithm focuses on parts of the search space that seem promising or uncertain. %This flexibility is especially important in systems where safety is critical and computational resources might be limited.

\subsection{Stability Analysis}

We could also analyze stability to learn more about the convergence and stability of our joint optimization framework. The insights from Theorem 1 and Theorem 2, which deal with Lipschitz continuity and convergence properties, help us understand the stability of our proposed approach. With these insights, we can study the stability of both the inner and outer loop optimization processes under different conditions, such as changes in fidelity settings and different environment configurations. In the end, this analysis helps us create algorithms that are more robust against uncertainties.

\subsection{Robustness Analysis}

Robustness analysis is about evaluating how our joint optimization framework performs when faced with varying levels of uncertainty and environmental noise, both in terms of configurations and simulator dynamics. By studying how our framework behaves under these conditions, we can pinpoint potential vulnerabilities to bolster its robustness. To carry out this analysis, we assess the impact of noise and uncertainty on the performance of both the inner and outer loop optimization processes. This might involve deriving robustness bounds or establishing worst-case performance guarantees, as well as exploring how fidelity settings affect the sensitivity of the optimization process to noise and uncertainty. Through comprehensive robustness analysis, we can confidently assert that our proposed approach is well-equipped to handle uncertainties in environment configurations and simulator dynamics.

\section{Conclusions}

We presented a mathematical formulation for joint falsification and fidelity setting optimization, which addresses the challenge of efficiently validating the safety of autonomous systems. The proposed framework brings together the two critical aspects of the problem, namely, the falsification of learning-enabled systems and the optimization of simulator fidelity settings. Our approach enables a more efficient exploration strategy for searching potential failure scenarios by focusing on the most critical environmental configurations that challenge the control algorithms. We have derived a set of six key theorems to establish the fundamental properties and relationships in the joint optimization problem. These theorems encompass a range of important aspects, including sensitivity analysis, sample complexity, convergence, the interplay between the outer and inner loops, and the regret bound analysis when employing Bayesian optimization. The insights gained from these theorems provide a foundation for the development of efficient algorithms in this domain. As a future direction, we aim to conduct extensive empirical evaluations of our approach on various autonomous systems to demonstrate its practical applicability and effectiveness in improving safety validation.

\bibliographystyle{splncs04}
\bibliography{ref}

\end{document}